\documentstyle[12pt,psfig,epsfig]{article}
\newcommand{\seta}{\rightarrow}

\newcommand{\beq}{\begin{equation}}
\newcommand{\eeq}{\end{equation}}

\begin{document}

\title{Vector Meson Production in the Golec-Biernat W\"usthoff Model} 
\author{Allen C. Caldwell \\ Columbia University,
New York, New York, USA\\
Mara S. Soares \\ Unicamp, Campinas, SP, Brazil}
\maketitle  

\abstract{We apply the Golec-Biernat W\"usthoff model
in the calculation of vector meson photo- and electroproduction. 
Starting from very simple non-relativistic wave functions we show 
that the model provides a good description of $J/\Psi$ cross 
sections in wide $Q^2$ and $W$ ranges. 
For the light mesons one obtains the approximately correct $W$ 
dependence and ratio of longitudinal to transverse cross sections, 
although in this case the 
normalization,
affected mainly by the wave function employed, is not in good agreement 
with data.}

\section{Introduction}

We consider elastic
vector meson (VM) production in $\gamma^*p$ scattering in the
context of the Golec-Biernat W\"usthoff (GBW) 
model~\cite{kgb_fit,kgb_dif}.
The model is constructed in the proton rest frame, as shown
in Fig.~1, where the  $q\bar{q}$ 
pair resulting from a photon fluctuation
can have a long enough lifetime to allow its interaction with the proton.
In this frame the $q\bar{q}$ longitudinal momentum is quite large
so that $\vec{r}$, the bi-dimensional separation of the pair, doesn't
 change  significantly during the scattering.
It is then natural to perform the calculations
in the  $(\vec{r},z)$ representation.

\begin{figure}[h]
\begin{center}
\psfig{figure=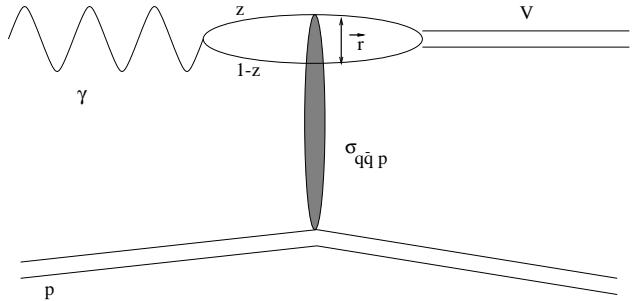,height=4.cm}
\end{center}
\caption{\it Scheme for vector meson production in the Color Dipole model.}
\end{figure}

The dynamics of the interaction between the $q\bar{q}$ pair and the
proton are absorbed in the dipole cross section, $\sigma_{q\bar{q}-p}$.
This approach to describing the scattering has been followed by many
authors (see for example~\cite{Mueller,Nikzak1,Nikzak2,BFGMS,GLM}), 
and is often called the Color Dipole model.
In the version proposed by K. Golec-Biernat \& M. W\"usthoff, 
the pair-proton cross section assumes the form
\beq
\sigma_{q\bar{q}-p}(Q,W,r)=\sigma_0\left(1-\exp\left[-\frac{r^2}{4R^2_0(Q^2,
W^2)}\right]\right), \ \ \ \ \ \ {\rm where}
\eeq
\begin{equation}
\label{eq:gbw}
R_0(Q^2,W^2)=\frac{1}{{\rm{GeV}}}\left[\frac{1}{x_0}\frac{Q^2+4m_q^2}{Q^2+W^2}
\right]^{\lambda/2}.
\end{equation}
One important characteristic of this parameterization is that the 
cross section
does not increase indefinitely with $r$, but ``saturates''  when
$r$ increases beyond $R_0$. $R_0$ therefore defines a saturation
radius, which depends on $Q^2$,  $W^2$, and the mass of the quark/antiquark.
The $W$ dependence of this cross section depends on the
ratio $r/R_0$:  for values of the ratio $\geq 1$, the $W$ dependence is
approximately flat, while for small values, $\sigma_{q\bar{q}-p}$  scales
approximately as $W^{2\lambda}$.  We therefore expect
a steeper $W$-dependence for heavy vector mesons relative to light
vector mesons at a given $Q^2, W^2$.  

The three free parameters 
($\sigma_0$, $\lambda$, $x_0$) in the expression above were 
determined via fits to DIS ($F_2$) data~\cite{kgb_fit}. 
The total $\gamma^*p$ cross
section is calculated as~\cite{ForshawRoss}:
\beq
\sigma_{L,T}(Q,W)=\int d^2\vec{r} \int dz |\Psi_{L,T}(Q,z,r)|^2
\sigma_{q\bar{q}-p}(Q,W,r)
\eeq
where $\Psi_{L,T}(Q,z,r)$ stands for the $q\bar{q}$ component of
the wave function of a 
longitudinally (L) or transversely (T) polarized photon. 
The proton structure function $F_2$, 
the data from which is used in the extraction of the parameters, 
is related to $\sigma_{L,T}$ via
\beq
F_2(Q^2,W^2)=\frac{Q^2}{4\pi^2 \alpha}(\sigma_{T}+\epsilon\cdot\sigma_{L}).
\eeq
The numbers resulting from the fit are  
$\sigma_0$ = 23.03 mb, $\lambda$ = 0.288 and
$x_0$ = 3.04$\times 10^{-4}$ if only three light quarks are assumed for
the photon wavefunction, and
$\sigma_0$ = 29.12 mb, $\lambda$ = 0.277 and $x_0$ = 0.41$\times 10^{-4}$ 
if one also considers charm. 

In these fits the light quark ($u,d,s$) mass is used as a parameter
and fixed to $m_{u,d,s}=140$ MeV; this average value was necessary to allow 
the model to give rough agreement with the total 
photoproduction cross section.  The charm quark mass used in
the calculations is discussed below.

The model has been applied in
the calculation of other inclusive processes, and has provided 
good results with no extra parameters. For example, it gives the
qualitatively correct behavior for the quantity $dF_2/d\ln{Q^2}$ 
at low $Q^2$. Furthermore the diffractive cross section $\sigma^{diff}$ is
reasonably reproduced in the model~\cite{kgb_dif}.

Our aim in this work is to test the model for exclusive processes. We
analyze the model in photo- and electroproduction of $J/\Psi,\ \rho$ and 
$\phi$ mesons. We first describe the calculations, and then compare the
results with published HERA data.

\section{Vector Meson Production}

In this section we give the principal formulae for the process
$\gamma^{(*)} p \seta Vp$ in the context of Color Dipole model. 
One writes the amplitude for the process as 
 $A =\ <V\ |\ \sigma_{q\bar{q}-p}\
|\ \gamma^{(*)}>$, where $|V>$ is the vector meson wavefunction and
$|\gamma^*>$ is the $q\bar{q}$ component of the $\gamma^*$ 
wavefunction.
In the $(\vec{r},z)$ representation $A$ 
is given by~\cite{Nikzak3}
\begin{equation}
A_{L,T} = \int_0^1 dz \int d^2\vec{r}\ \Psi_{V}^{* L,T}(r,z)\ 
\sigma_{q\bar{q}-p}(Q,W,r)\
\Psi_{\gamma}^{L,T}(Q,r,z)
\label{amplit}
\end{equation}
where 
$\sigma_{q\bar{q}-p}(Q,W,r)$ is the cross section for the pair-proton
interaction. The forward scattering cross section is then given by
\beq
 \frac{d\sigma}{dt}|_{t=0}
 = \frac{\ |A|^2}{16\pi}.
\eeq
Since the model does not incorporate any $t$ dependence we assume 
the ordinary exponential $t$ dependence, $d\sigma/dt \simeq C e^{-B|t|}$
observed in the data,
 so that the total cross section is given by
\begin{equation}
\label{sigmaVp}
\sigma(\gamma p \seta V p) = \frac{1}{B}
\frac{d\sigma}{dt}| _{t=0} = \frac{1}{B}
\frac{|A|^2}{16 \pi}.
\end{equation}

The $q\bar{q}$ component of the
photon wave function is quite well known~\cite{photon}. For 
photons with longitudinal polarization we have
\begin{equation}
\Psi_\gamma^L=\frac {\sqrt{N_c}}{(2 \pi)^{3/2}} \sqrt{4 \pi \alpha}
\ e_q\ 2Qz(1-z)\phi^\gamma(z,r)\delta_{\lambda_1,-\lambda_2}
\end{equation}
and for transverse polarization
\begin{eqnarray}
\nonumber
\Psi_\gamma^T=\frac {\sqrt{N_c}}{(2 \pi)^{3/2}} \sqrt{4 \pi \alpha}
\ e_q \{m_q \left( \begin{array}{l} \mp 1\\
-i \end{array} \right) \phi^\gamma(r,z)\delta_{\lambda_1,\lambda_2}
+ \\
\left( \begin{array}{l}i(2z-1)\hat{b}_x \mp \hat{b}_y \\
\pm \hat{b}_x+i(2z-1)\hat{b}_y \end{array} \right)\frac{\partial
\phi^\gamma(r,z)}{\partial r}\delta_{\lambda_1,-\lambda_2} \}.
\end{eqnarray}
In the expressions above $N_c$ is the number of colors,
$e_q (m_q$) is the  quark charge (mass),
$\lambda_{1,2}$ are the helicities of the $q$ and $\bar{q}$, and
$\phi^\gamma (r,z)$ is the photon spatial wave function given by
\beq
\phi^\gamma (r,z)=K_0(r\sqrt{Q^2z(1-z)+m_q^2}\ ).
\eeq

In analogy with the photon wave function one can write for the 
longitudinally polarized vector mesons
\beq
\Psi_V^L=-2M_V\phi^V(z,r)\delta_{\lambda_1,-\lambda_2}
\eeq
and for the transversely polarized vector mesons
\begin{eqnarray}
\nonumber
\Psi_V^T=\frac{m_q}{z(1-z)}\left( \begin{array}{l} \pm 1\\
-i \end{array} \right)\phi^V(z,r)\delta_{\lambda_1,\lambda_2} + \\
\frac{1}{z(1-z)}\left( \begin{array}{l}i(2z-1)\hat{b}_x \mp \hat{b}_y \\
\pm \hat{b}_x+i(2z-1)\hat{b}_y \end{array} \right)\frac{\partial
\phi^V(r,z)}{\partial r}\delta_{\lambda_1,-\lambda_2} .
\end{eqnarray}

In the case of the vector mesons, the spatial part of the wave function is not 
known, and some hypothesis must be made.
In the following sections we
discuss the vector meson wave functions.

\section{$J/\Psi$ wave function}

Since the internal motion is slow for heavy quarkonium, we can choose
a non-relativistic wave function. 
For transversely polarized photons, $J_z=\pm 1$. If we assume that the orbital
angular momentum of the quark and antiquark in the $J/\psi$
is 0, then $S_z=\pm 1$.  This implies 
$\lambda_1 =  \lambda_2$, and the second term in Eq. (12) 
disappears (see also ~\cite{MartinWuesthoff}).
We analyze some possible wave functions in $(z,k_t)$ space, and then
calculate the $(z,r)$ wavefunction using a 2-D Fourier transform,
\begin{equation}
\label{fourier}
\phi_V(r,z) = \int \frac{d^{2}k_t}{ 4\pi^2 }\ \phi_{V}(z,k_t)
e^{ir\cdot k_t}\;\; .
\end{equation}

\subsection{$\delta$-function}

The simplest case one can have is to consider that the $q$ and $\bar{q}$ have 
the same 
longitudinal momentum fraction  and that the transverse momentum is 
very small.
In terms of a $(z,k_t)$ space wavefunction, this hypothesis yields
\begin{equation}
\phi_{V}(z,k_t)=K\ \delta(z-1/2)\delta^2(k_t)\; \; .
\label{delta}
\end{equation}
The only free parameter is the normalization, $K$. It can be determined 
if we fix the partial width for $J/\psi \rightarrow e^+e^-$ to the
experimentally measured value.
\begin{equation}
\Gamma^{V}_{e^+e^-}=\frac{32 \pi\alpha^2 e_q^2}{M_V}\ | \int dz \int \frac{
d^{2}k_t}{8\pi^{3/2}} \Psi_V (z,k_t) |^2
\end{equation}
resulting in
\begin{equation}
K=\frac{1}{2 M_V} \frac{8\pi^{3/2}}{e_q \alpha}\sqrt{\frac{\Gamma^{V}_{e^+e^-}
M_V}{32 \pi}}.
\end{equation}
The wave function in the $(r,z)$ space representation, obtained via
Eq. (\ref{fourier}), is then
\begin{equation}
\phi_V(r,z) = \frac{1}{2 M_V} \frac{\sqrt{4 \pi}}{e_q
\alpha}\sqrt{\frac{\Gamma^{V}_{e^+e^-} M_V}{32 \pi}} \delta(z-1/2).
\end{equation}
The full expression for longitudinal and transverse wave functions
is obtained by inserting the expression above in Eqs.~(11) and (12).

\subsection{Gaussian function}

Another possibility is to consider again the same longitudinal 
momentum fraction 
for both $q$ and $\bar{q}$ but now we assume that some small
transverse momentum
is allowed inside the $J/\Psi$. Following~\cite{gauss},
we use a Gaussian distribution around zero and
the wave function reads
\begin{equation}
\phi_{V}(z,k_t)=K\ \delta(z-1/2)\exp(- a \frac{k^2_t}{m_c^2}).
\end{equation}
We take for $a$ the value obtained from lattice QCD calculations, 
$a=4.68$ and $m_c=$1.43 GeV.

Again the normalization is obtained from the decay width,
\begin{equation}
K=\frac{a}{m_c^2}\frac{1}{2 M_V } 
\frac{8\pi^{3/2}}{e_q \alpha}\sqrt{\frac{\Gamma^{V}_{e^+e^-}
M_V}{32 \pi}}
\end{equation}
and the $(r,z)$ representation of the wave function comes from the Fourier 
transform.

\subsection{Double-Gaussian function}

We propose now that not only the transverse momentum may have
some distribution but also the
longitudinal momentum fraction has a Gaussian distribution around 1/2,
resulting in
\begin{equation}
\phi_{V}(z,k_t)=K\ \exp[-c^2(z-1/2)^2]\exp(- a \frac{k^2_t}{m_c^2}).
\end{equation}
Now the wave function has one extra parameter, $c$.
However, one extra constraint is available since now the wave function 
can be normalized.  From the decay width
\begin{equation}
K=\frac{a}{m_c^2}\frac{1}{2 M_V } \frac{8\pi^{3/2}}{e_q \alpha}
\sqrt{\frac{\Gamma^{V}_{e^+e^-}
M_V}{32 \pi}}\frac{c}{\sqrt{\pi}\ {\rm erf}(c/2)},
\end{equation}
and from the normalization condition,
\beq
\frac{N_c}{2\pi}\int_0^1\frac{dz}{z^2(1-z)^2} \int d^2\vec{r}
\{ m_c^2|\phi(r,z)|^2 \}
=1
\eeq
we get $c^2=27.2$. In this way the $(r,z)$ representation of
the wave function is obtained via the Fourier transform with no
free parameters.

\section{Light meson wave functions}

For the light mesons, a relativistic approach must be used.
We follow the procedure outlined in~\cite{fks,predazzi}, where
one writes the wave function in terms of the lightcone invariant
variable $\vec{p}^{\ 2}$ (3-momentum of the quark in the non-relativistic
limit),
\beq
\vec{p}^{\ 2} = \frac{1}{4}(M^2-4 m_q^2),\ \ \ \ \ \ \ \ \ 
M^2=\frac{m_q^2+k_t^2}{z(1-z)}.
\eeq
In this expression $M$ is the invariant mass of the $q\bar{q}$ system,
and $m_q=140$~MeV is the quark mass.
Again a Gaussian form is applied, now in the $\vec{p}$ variable,
\beq
\phi(\vec{p})= K\ (2\pi R^2)^{3/2}\exp\left[-\frac{1}{2}p^2R^2\right].
\eeq
The value of $R$ was fixed by requiring that the exponential gives a
value of $1/e$ for $M=M_V$.  This yielded

\centerline{$R^2_\rho=15.5$ GeV$^{-2}$,}

\centerline{$R^2_\phi=8.3$ GeV$^{-2}$.}
\noindent Using Eqs. (23)  above one obtains the wave function in
the $(z,k_t)$ representation, and through Eq. (13) 
\begin{equation}
\phi(r,z)=K\  4z(1-z)\sqrt{2\pi R^2}
\exp[-\frac{m_q^2 R^2}{8z(1-z)}] 
\exp[-\frac{2z(1-z)r^2}{R^2}]\exp[-\frac{m_q^2 R^2}{2}] \; \; .
\end{equation}
The normalization factor $K$ is determined from the normalization condition,
so that once again there are no free parameters.


\section{Results}

We have fixed the $u,d,s$ quark masses to 140 MeV, as in~\cite{kgb_fit}, and
use either the three or four quark parameters given in the introduction
(for the $J/\psi$ case, only the four quark parameters are used).
For the slope $B$ (see Eq. (\ref{sigmaVp})) we use a parameterization
to experimental data, given by ~\cite{BruceMellado}
\beq
B = 0.60 \left(\frac{14}{(Q^2+M_V^2)^{0.26}} + 1\right) \; \; .
\eeq
This parameterization is used for all the vector mesons.

\subsection{$J/\Psi$ production}

In Fig. 2,  the $W$ dependence of the $\gamma^*p$ cross section for 
$J/\Psi$ elastic production is shown for various values of $Q^2$.
It can be seen that even for the simplest wave function utilized 
($\delta$-function, dotted line) the $W$ dependence predicted by the model 
is in agreement with the data,
although for photoproduction the normalization is not very good for
this wave function. For the more complicated forms we obtain
good agreement to HERA data both in shape and in normalization.
Note that the value of $R_0$ in Eq.~(\ref{eq:gbw}) is never very small
in $J/\psi$ production because of the presence of the charm quark mass,
implying that we are far from the low $Q^2$ or high $W$ saturation regions.  
The $W$ dependence is then given to good approximation by 
$\sigma_{\gamma^*p\rightarrow J/\psi p} \propto W^{4\lambda}$.
\begin{figure}[htb]
\begin{center}
\psfig{figure=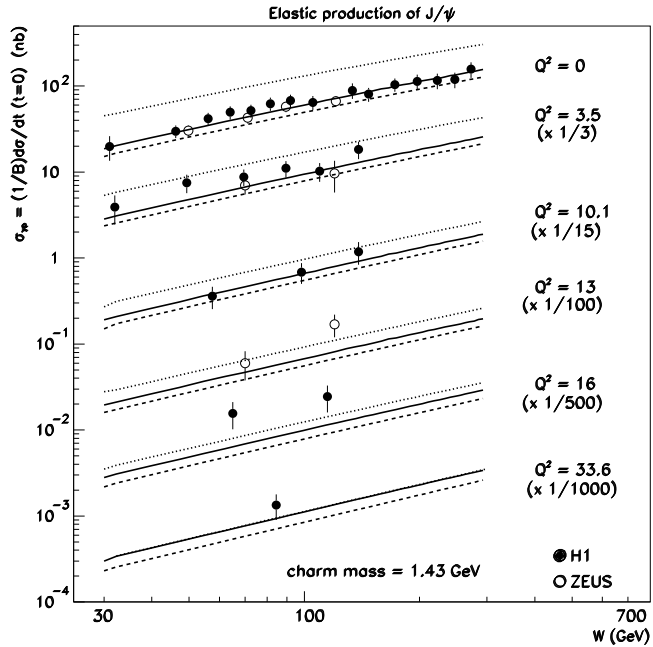,height=10.cm}
\end{center}
\caption{\it The W-dependence of elastic $J/\Psi$ production as a function
of $Q^2$. Open circles are 
ZEUS data~\cite{ZEUSpsi1,ZEUSpsi2}
and solid circles H1 data~\cite{H1psi1,H1psi2}. The dotted lines correspond
to calculations using the $\delta$-function wave function, 
the dashed lines to calculations using a single Gaussian, 
and the solid lines to the double Gaussian wavefunction as discussed
in the text. The $Q^2$ values, in units of GeV$^2$, are 
indicated in the figure.  The data and calculations have been scaled
by the values given for clarity of presentation.}
\end{figure}

Fig.~3 gives the ratio of the longitudinal to transverse cross sections,
$\sigma_L/\sigma_T$, for fixed $W$ as a function of $Q^2$ (upper
plot), and for fixed $Q^2$ as a function of $W$ (lower plot). 
Due to the delta function in $z$ for the $\delta$ and Gaussian wave functions,
the integration
over $d^2r dz$ is the same for transverse and longitudinal amplitudes, up
to a constant factor. In this case the model
dependence for $R$ disappears and one gets just
$R=Q^2 M_V^2/(16 m_q^4)$.  The dotted and dashed curves are therefore
superimposed in Fig.~3. For the double Gaussian case this does not occur;
$A_L$ is peaked near $z=1/2$ and the $Q^2$ growth is damped relative to
$A_T$.  It is clear that the $Q^2$ dependence of $R=\sigma_L/\sigma_T$
is quite sensitive to the vector meson wavefunction, while there is
little sensitivity in the $W$ dependence. The data favor the
double Gaussian wavefunction.

\begin{figure}[htb]
\begin{center}
\psfig{figure=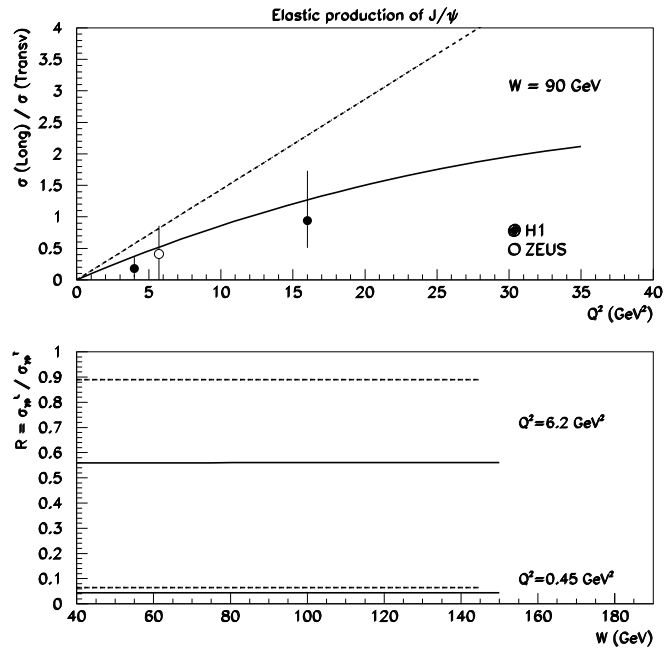,height=10.cm} 
\end{center}
\caption{\it Ratio $R=\sigma_L/\sigma_T$ for elastic $J/\Psi$ 
electroproduction. The
dotted and dashed lines (superimposed) correspond to the $\delta$ and 
single Gaussian wave functions. The solid lines are calculated with 
Double Gaussian wave function. The upper plot gives the $Q^2$
dependence at fixed $W=90$~GeV.  The lower plot gives results versus
$W$ for $Q^2$ = 6.2 GeV$^2$ (upper curves),
 and $Q^2$ = 0.45 GeV$^2$ (lower curves).
The open circles are ZEUS data~\cite{ZEUSpsi2} and the solid
circles are H1 data~\cite{H1psi2}.}
\end{figure}

\subsection{$\rho$ production}

The comparisons to data for elastic $\rho$ production are given in
Figs.~4-6.  The lighter mass of the $\rho$ allows the
 cross sections to enter the saturation region ($r \sim R_0$), 
and the $W$ dependence is a strong function of $Q^2$, as seen in Fig.~4.  
The two sets of curves on the
graph give the results for three quark (dashed) and four
quark (solid) parameter sets.
The normalization of the cross sections is not particularly good, and
the $Q^2$ dependence is also not well reproduced, although the 3 parameter
set does a better job of reproducing the $Q^2$ dependence. Note that the
normalization is
affected both by the choice of $\rho$ wavefunction and by the 
assumed $B$ values.  However, the observed change of the $W$
dependence is rather well reproduced.  This is quantified in Fig.~5,
where the power $\delta$ from a fit of the $W$ dependence with a form
$\sigma_{\gamma^*p \rightarrow \rho p} \propto W^{\delta}$ is compared with
data.

\begin{figure}[htb]
\begin{center}
\psfig{figure=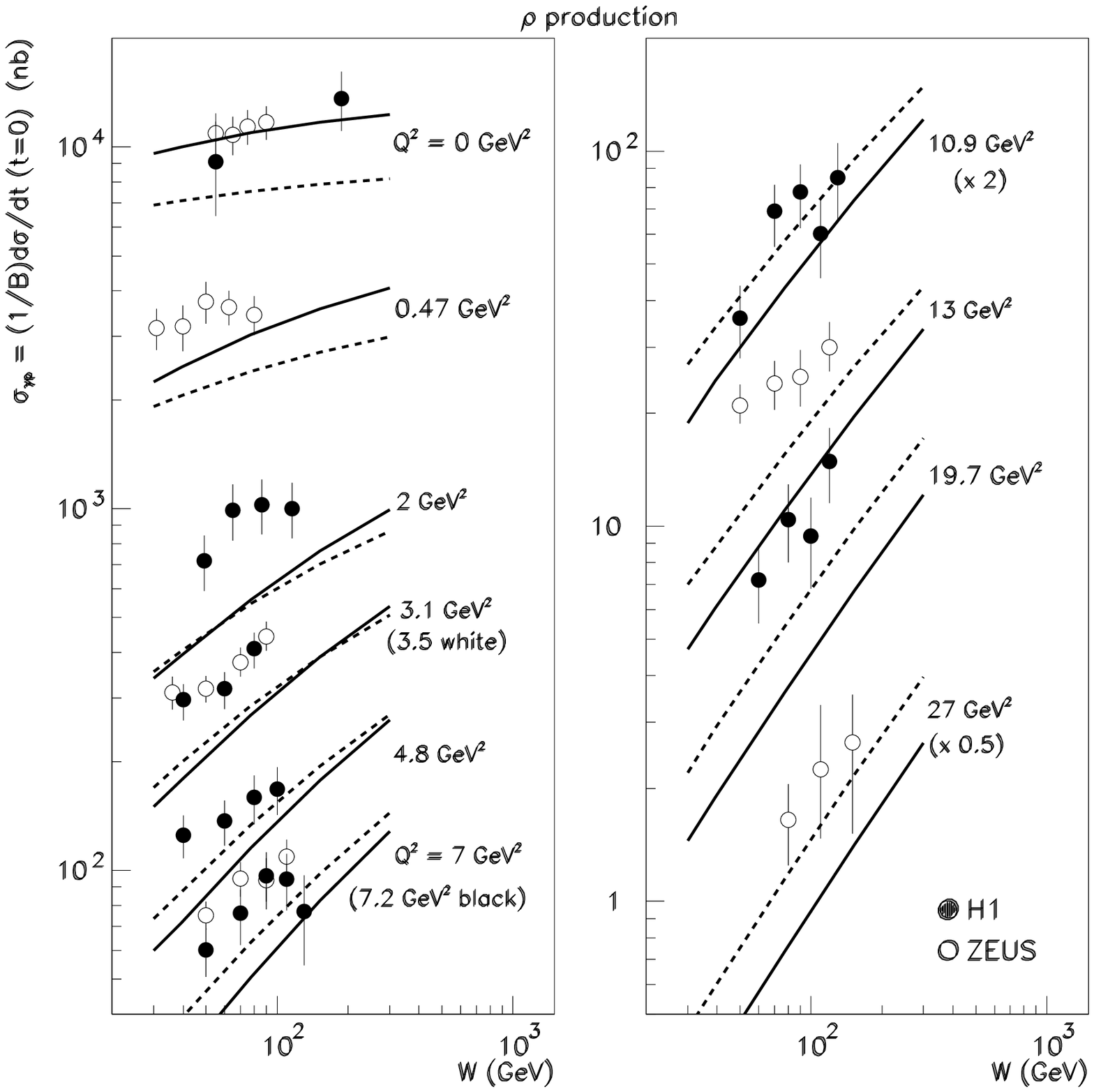,height=12.cm}
\end{center}
\caption{\it The $W$-dependence of elastic $\rho$ production as a function
of $Q^2$  compared with recent ZEUS~\cite{ZEUSrho1,ZEUSpsi2} and 
H1~\cite{H1rho1,H1rho2} data.  The $Q^2$
values are given in the plot. The solid lines are for the four quark 
parameters, while the dashed lines are for the three quark parameter sets.}
\end{figure}

\begin{figure}[htb]
\begin{center}
\psfig{figure=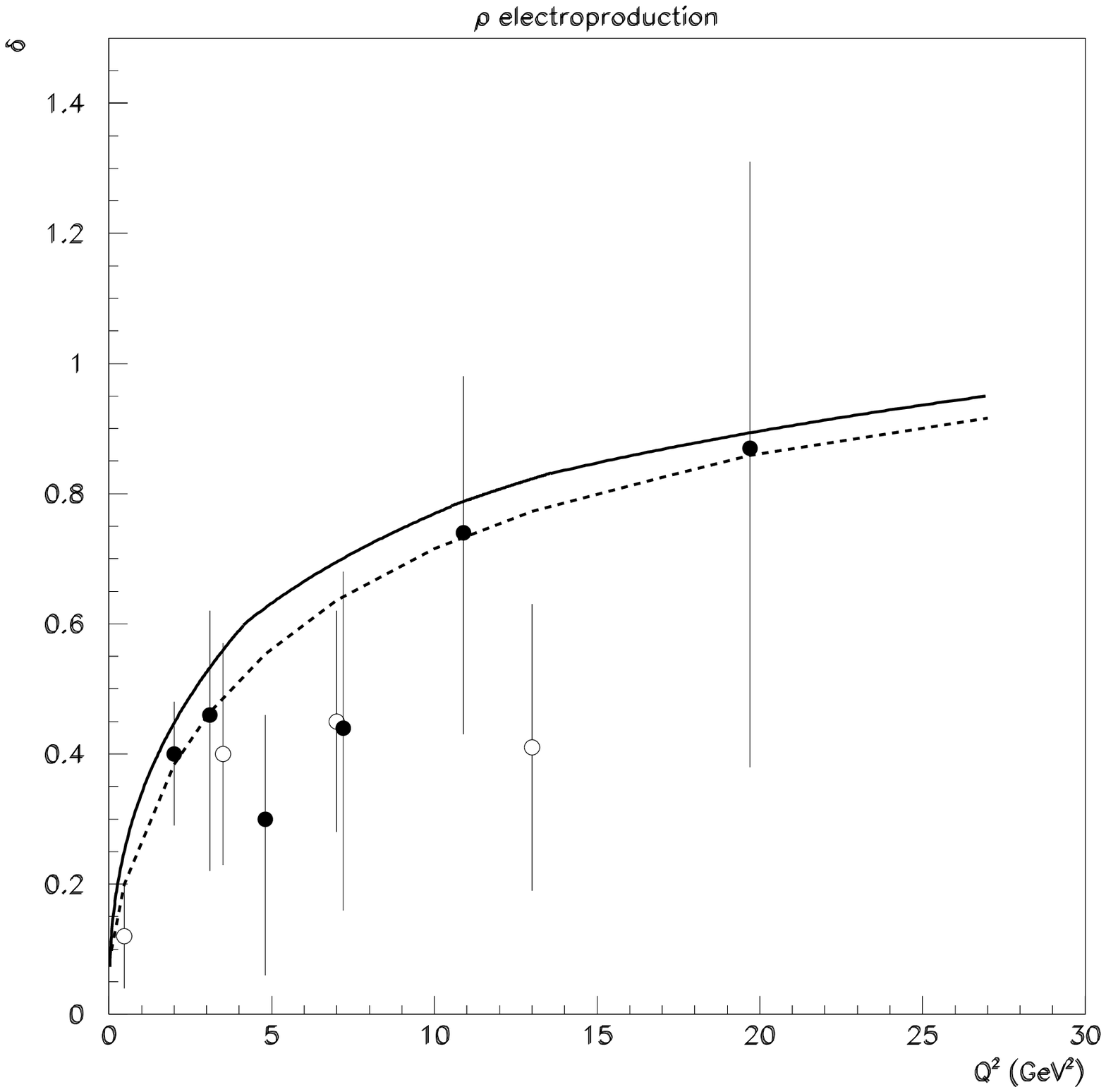,height=8.cm}
\end{center}
\caption{\it The $W$-dependence of elastic $\rho$ production as a function
of $Q^2$, compared with recent ZEUS~\cite{ZEUSpsi2} (open circles)
and H1~\cite{H1rho2} (solid circles)
data.  The solid line is the result using the four flavor parameter set, and
the dashed curve is the result using the three flavor parameter set.}
\end{figure}
    
The dependence of $R=\sigma_L/\sigma_T$ 
on $Q^2$ at fixed $W$ is shown in Fig.~6, as
is the $W$-dependence at fixed $Q^2$.  The model, with the simple
wavefunction chosen, gives a good representation of the data.  
In this
case, the model predicts a small variation of $R$ with $W$.
 
\begin{figure}[htb]
\begin{center}
\psfig{figure=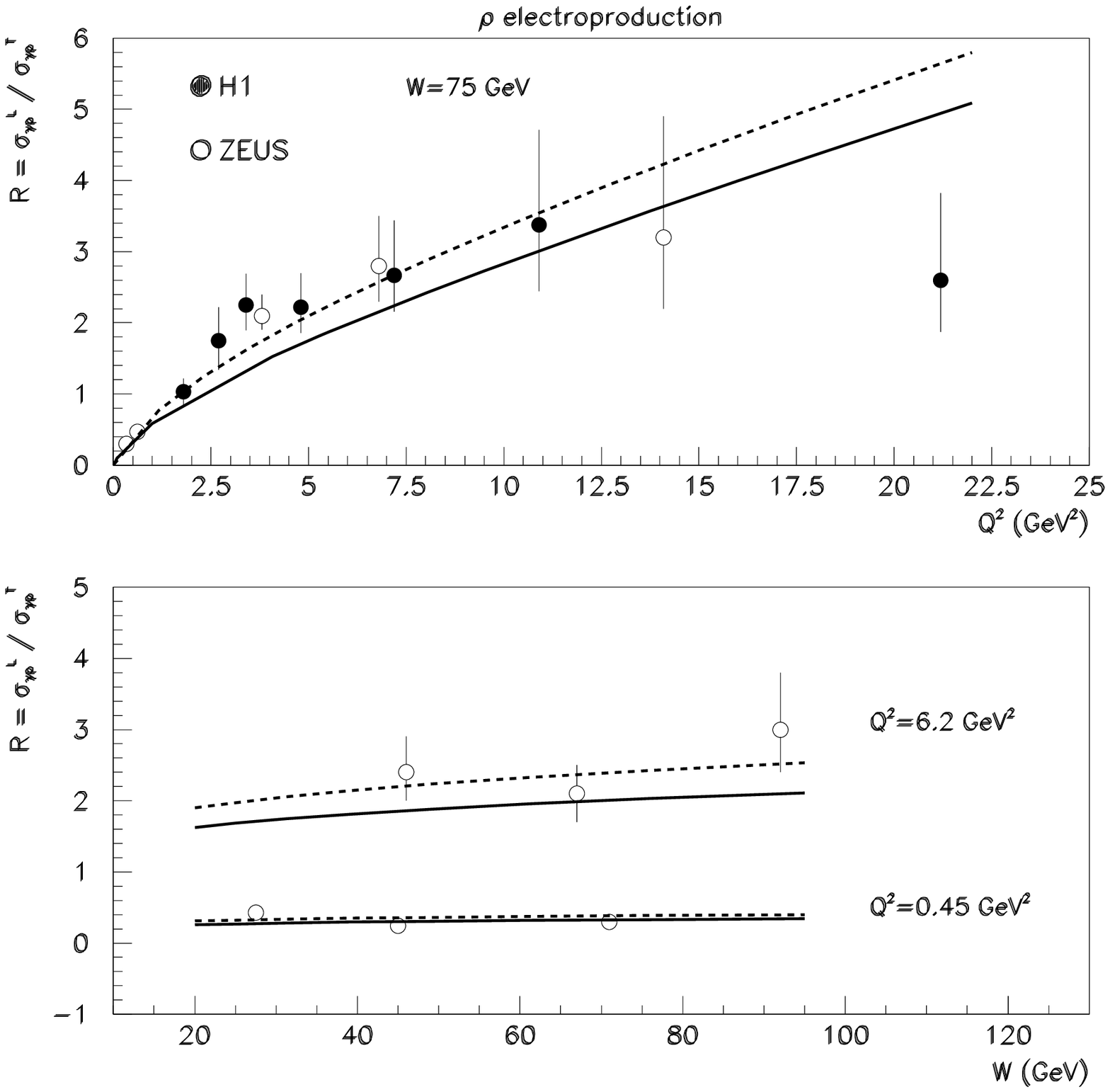,height=12.cm}
\end{center}
\caption{\it $R$ dependence on $Q^2$ and $W$ in 
elastic $\rho$ production. The upper plot gives the $Q^2$ dependence at
fixed $W=75$~GeV, while the lower plot gives the $W$ dependence at the $Q^2$
values indicated in the plot. The open circles are 
ZEUS data~\cite{ZEUSpsi2}
and the solid circles H1 data~\cite{H1rho2}.}
\end{figure}

\subsection{$\phi$ production}

For the $\phi$ meson, we compare the predicted $W$ dependence for
$\gamma^*p \rightarrow \phi p$ as a function of $Q^2$ with
data in Fig.~7 and $R$ versus $Q^2$ at fixed $W$, as well as
$R$ versus $W$ at fixed $Q^2$ in Fig.~8.  The normalization
of the cross section is too high given the chosen wavefunction
and parameterization for $B$, but the $R$ dependence on $Q^2$ is in
reasonable agreement with the available data.  

\begin{figure}[htb]
\begin{center}
\psfig{figure=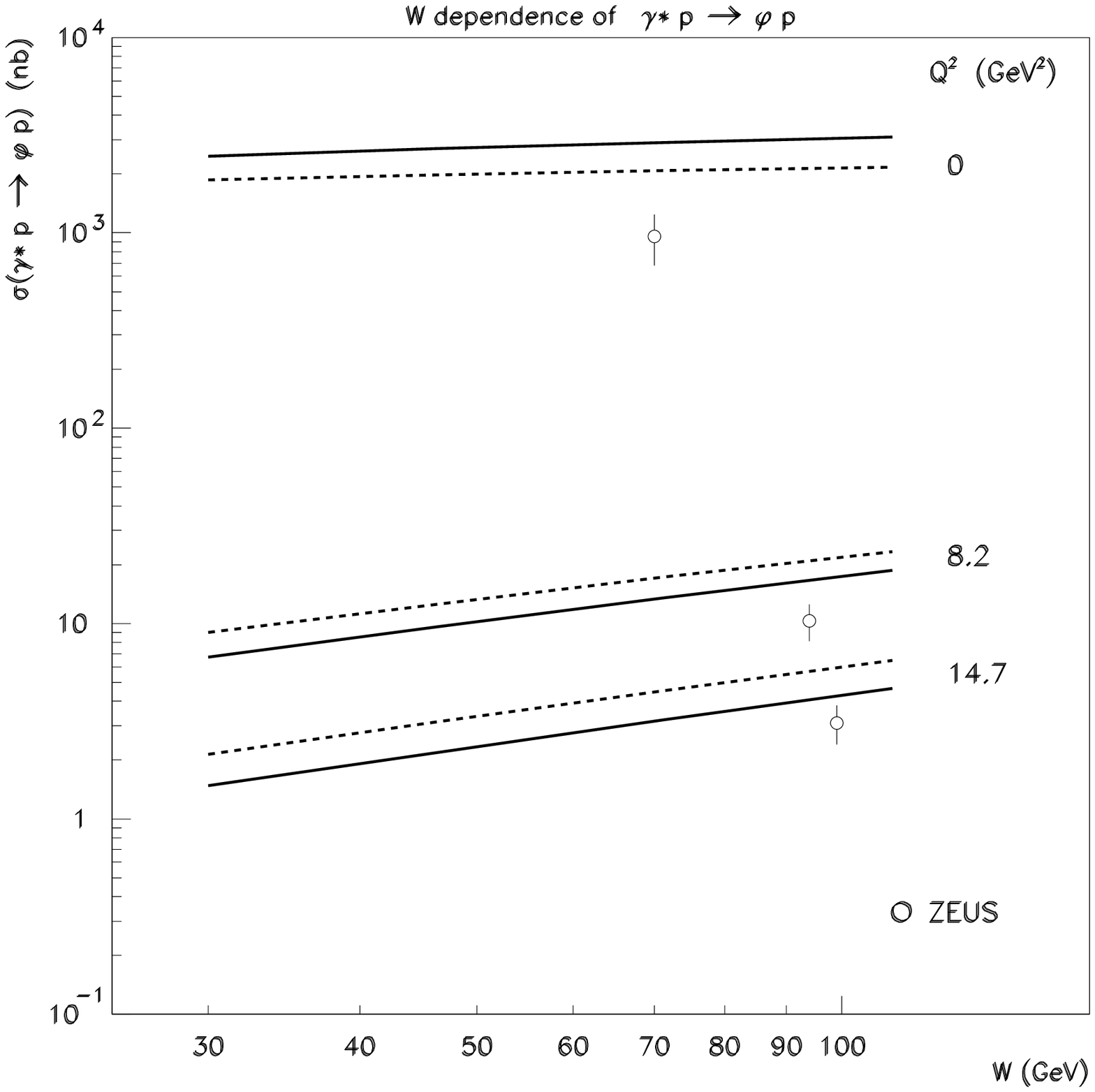,height=10.cm}
\end{center}
\caption{\it $\sigma_{\gamma^*p \rightarrow \phi p}$ as a function
of $W$ for different $Q^2$, compared with ZEUS data~\cite{ZEUSphi}.
The solid lines are for the four flavor parameter set, while the
dashed lines are for the three flavor parameter set.}
\end{figure}

\begin{figure}[htb]
\begin{center}
\psfig{figure=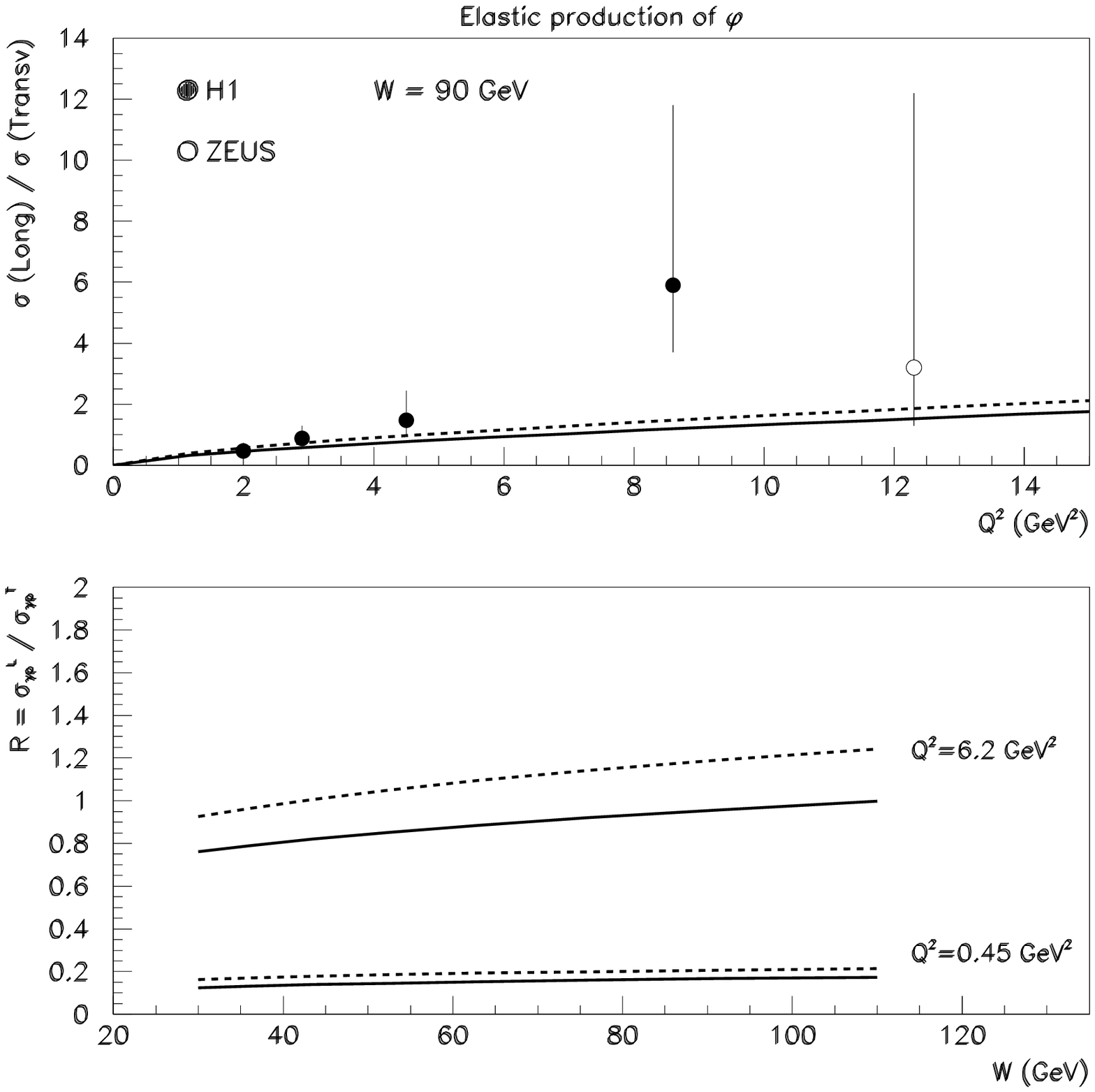,height=12.cm} 
\end{center}
\caption{\it Ratio of longitudinal over transverse cross sections in 
$\phi$ production. The upper plot shows the $Q^2$ dependence at
fixed $W$, while the lower plot shows the $W$ dependence at fixed 
$Q^2$. The solid lines are for the four flavor parameter set, while the
dashed lines are for the three flavor parameter set.  The data are from
ZEUS~\cite{ZEUSphi} and H1~\cite{H1phi}.}
\end{figure}

\section{Summary and Conclusions}

We have used the Golec-Biernat W\"usthoff model to calculate
the cross section for exclusive vector meson production in $ep$
scattering.
The model gives the correct $W$ dependence for both photo- and
electroproduction of $J/\psi$ mesons. 
This feature can be attributed to the
model because it is observed even when a very simple wave function is
employed.  This was not unexpected, since the model reproduces the
pQCD expectations for the $Q^2$ and $W$ dependence of 
$\sigma_{\gamma^*p \rightarrow \psi p}$
when the dipole size
is small compared to the ``saturation radius''.  Perhaps more surprising
is the fact that a simple $J/\psi$ wave function (double Gaussian)
also gives good results for the normalization of the wavefunction 
as well as for the $Q^2$ dependence of $R=\sigma_L/\sigma_T$.

For $\rho$ and $\phi$ production, the $W$ dependence 
changes dramatically as we go from photoproduction to DIS.  The change in
the $W$ dependence versus $Q^2$ calculated using a simple wavefunction
can reproduce the data reasonably well.  
The behavior of $R$ versus $Q^2$ and $W$ is
also in reasonable agreement with the data.  However, the normalization of the
cross section is not in agreement with the data, and the discrepancy with data
varies with $Q^2$.  This problem can be attributed to
the lack of knowledge of the wave function and perhaps also in part to the
change in $t$-slope with $Q^2$ (this slope is needed to calculate the
total cross section).

The proton rest frame description of $ep$ scattering gives a simple
and intuitive picture of low $Q^2$ and small-$x$ scattering.  Golec-Biernat
and W\"usthoff have shown, using a three parameter parametrization for
the $q\bar{q}-p$ scattering cross section, that the total as well as the
diffractive cross sections can be well described.  We find that the
the model can also describe many of the features of exclusive vector
meson production.  

\centerline{\bf Acknowledgments}
The authors would like to thank the U.S. National Science Foundation and the
Brazilian agency FAPESP for support during the period of this study.
One of us, M.S., would also like to thank Columbia University for its 
hospitality during this period.  We are grateful for useful discussions 
with H. Abramowicz, Z. Chen, R. Covolan, J. Gravelis, H. Kowalski, 
R. Mawhinney, A. Mueller, J. Nemchik,
E. Predazzi, F. Sciulli, and M. Strikman.

\end{document}